\documentclass[usenatbib,twocolumn]{mn2e}
\usepackage{graphicx}
\usepackage{amssymb}

\usepackage[authoryear]{natbib}
\usepackage{subfigure}
\usepackage{graphicx,color}

\usepackage{amssymb}
\def\beq{\begin{equation}}
\def\eeq{\end{equation}}
\def\apj{ApJ}
\def\apjs{ApJS}

\def\ep{\epsilon}

\def\araa{ARAA}
\def\nat{Nature}
\def\mnras{MNRAS}
\def\apj{ApJ}
\def\apjl{ApJL}

\def\beq{\begin{equation}}
\def\ee{\end{equation}}
\def\lsim{\mathrel{\rlap{\lower4pt\hbox{\hskip1pt$\sim$}}
    \raise1pt\hbox{$<$}}}
\def\gsim{\mathrel{\rlap{\lower4pt\hbox{\hskip1pt$\sim$}}
    \raise1pt\hbox{$>$}}}

\def\ts{\times}

\def\B{{\bf B}}

\newcommand{\eric}[1]{\textcolor{blue}{[EGB: #1]}}

\title[Fragmentation Cascade and  Feedback]
{Cloud fragmentation cascades and feedback:  on reconciling an unfettered inertial range with a low star formation rate}

\author [Blackman]
{Eric G. Blackman$^{1}$\thanks{E-mail: blackman@pas.rochester.edu}
\\
 $^{1}$Department of Physics and Astronomy, University of Rochester, Rochester NY, 14627, USA\\
 }
\begin{document}
\date{}
\pagerange{\pageref{firstpage}--\pageref{lastpage}} \pubyear{}
\maketitle
\label{firstpage}
\begin{abstract}
Molecular cloud complexes exhibit both (i) an unfettered Larson-type 
 spectrum over much of their dynamic range, whilst (ii) still producing a much lower star-formation rate than were this cascade to remain unfettered all the way down to star-forming scales.
Here we explain the compatibility of these attributes with minimalist considerations  of a mass-conserving fragmentation cascade, combined with  estimates of stellar feedback. Of importance is that the amount of feedback needed to abate fragmentation and truncate the complex decreases with decreasing scale.  The scale at which the feedback momentum matches
the free-fall momentum marks a  transition scale below most of the cascade is truncated and  the molecular cloud complex dissipated.   For a $10^6M_\odot$ GMC complex starting with radius of $\sim 50$pc, 
the combined feedback from young stellar objects, supernovae,   radiation, and stellar winds for a GMC cloud complex   can truncate the cascade within an outer free-fall time but  only after the cascade reaches parsec scales.
 \end{abstract}


\section{Introduction}

Modeling molecular cloud structure and  observed star-formation rate (SFR) of the Galaxy
is a complex task, involving a nonlinear interplay between
self-gravity, turbulence and stellar feedback 
in molecular cloud evolution \citep[e.g][]{Matzner+2000,Krumholz2006,McKee+2007,Dobbs+2014,Krumholz+2014,Krumholz+2018,Kim+2019}.
The effort to identify simplifying principles can be  constructive.

The hierarchical structure of molecular clouds is  consistent 
with an origin from fewer large clouds subjected to turbulent perturbations that facilitate fragmentation   to smaller scales
\citep{Hoyle1953,Chieze1987,Field+2008,Hopkins2013}.
Predictions for  mass, number, and density of molecular clouds as a function of scale emerge from this framework.   As applied to the Galaxy, the  outer scale  corresponds to  the largest giant molecular clouds (GMC)  of radii $\gsim 50$pc that source fragmentation into a quasi-self similar complex of clouds down to $\sim 0.1$pc clump scales.  
Hereafter, ``cloud complex (CC)" refers to the  entire  structure of subunits that emerges from an outer scale cloud.


The  connection between a fragmentation cascade and  the need for stellar feedback  is as
 follows: the total mass of molecular gas in the ISM is approximately
$2\times 10^9M_\odot$. Most of this is contained in $\sim 2000$ of the most massive GMC
\citep{Combes1991},   implying a mass $M_L\sim 10^6M_\odot$ per complex.  
Taking the characteristic radius of  these largest clouds to be $R_L\sim 50$pc, the  average mass density 
is then $\rho_L\sim 1.6\ts 10^{22}{\rm g/ cm^3}$.
Without kinetic  support, the clouds would collapse on a free fall time $t_{ff,L}$, which for a homogeneous sphere,
is 
\beq 
t_{ff,L}=
\left({3\pi\over 32 G \rho_L}\right)^{1\over 2} 
%
\sim 5.5 \ts 10^6\left(\rho_L\over 1.6\ts 10^{-22}{\rm g/cm^3}\right)^{-{1\over 2}}\ {\rm yr}.
\label{free}
\ee
If star formation were 100\% efficient and occurred on  a time scale $t_{ff,L}$, then dividing  cloud mass by
this free fall time gives the unfettered  ``no-feedback" SFR \citep{Zuckerman+1974,McKee+2007,Federrath2015}.
For $ 2000$  of these large GMC,
the  total Galactic SFR is then 
\beq
\begin{array}{r}
{\dot M}_{nf,T}= N{\dot M}_{nf}\sim 360 
\left({N\over 2000}\right)\left({M_L\over 10^6{M_\odot}}\right)\\ \times \left(\rho_0\over 1.6\ts 10^{-22}{\rm g/cm^3}\right)^{-{1\over 2}}
M_\odot/{\rm yr},
\end{array}
\label{totalnf}
\ee
where $\dot M_{nf}$ is the no-feedback SFR per large GMC.
The
 observationally inferred Galactic SFR is
${\dot M}_{obs}\sim 4M_\odot/{\rm yr} \sim \dot M_{nf,T}/100$
\citep{vandenBergh1991,Stahler+2004,Diehl+2006}.


Normalizing  the  SFR  by  mass and free fall time at a given scale $R$, we can construct a  dimensionless ratio 
\beq
\epsilon_{ff}(R)= {{\dot M}_{obs} f_R  t_{ff}(R)\over M_R},
\label{3}
\ee
where $f_R$ is the fraction of star formation occurring within structures
of scale $R$, where   $R$  ranges from $R_L$  down  to clump
scales, $M_R$ is the sum of the mass in  all structures with scale $R$ (later made explicit in
equation (\ref{MR})), and
$t_{ff}$ is the free fall time  for structures of scale $R$.
Equation (\ref{3}) is inspired by equation (1) of  \cite{Krumholz+2007}:
\beq
SFT_{X-ff}= {{\dot M}_{obs} f_X  t_{ff,X}(R)\over M_X},
\label{4}
\ee
where $f_X$ is the fraction of galactic star formation occuring in objects of some specified class $X$,
 $M_X$ is  the total mass class  $X$ objects in the Galaxy, and $t_{ff,X}$ is their free fall time.
Were we to allow  ``objects of class $X$" to represent ``objects of scale $R$," then 
equations (\ref{3}) and (\ref{4}) could be the same.  \cite{Krumholz+2007} do not use  $X$
in equation  (\ref{4}) to primarily distinguish objects  by scale, but 
by observation type.  \cite{Krumholz+2014}  compiled and reviewed  data indicating that
$SFT_{X-ff}\simeq 0.01$,  independent of density of the object class. 
Inasmuch as objects of a given scale in the  cascade also correspond to objects
of a given density, there is some correspondence between equations (\ref{3}) and (\ref{4}).



Here  $\epsilon_{ff}$ can be immediately seen to be constant
for a mass conserving fragmentation cascade: 
 First, ${\dot M}_{obs}$ is  an  observed constant. Second, the ratio of ${t_{ff}(R)\over M_R}$ is 
just the inverse of an unfettered constant mass throughput through the cascade that reflects the mass conserving assumption.
Third, since the cloud complex forms a nested structure with smaller clouds   embedded in  larger clouds,  all star formation
that occurs within the nested cloud structure is contained in all larger clouds, 
making  $f_R$ in equation (\ref{3}) is also constant.   

We will discuss how, in the fragmentation cascade,
  a low SFR $\epsilon_{ff}<<1$ is  consistent
with unfettered fragmentation over most of the observed
"inertial range"  down to a transition scale.
This follows  because the feedback needed to abate
the cascade decreases with decreasing scale and because 
combined feedbacks primarily abate the cascade at small truncation scales.
There is cognizance of the potential influence of different feedbacks on different  scales  \citep[e.g][]{Kim+2019},
but here we pithily combine  a fragmentation cascade with  feedbacks   to estimate a  transition scale. 


 


In section 2, we  discuss the  conceptual picture and quantitative scalings of the fragmentation cascade, including the predicted number spectrum and size-velocity-density  relations. 
In section 3, we  show quantitatively that  the momentum (and  energy) of feedback needed to abate the fragmentation cascade and reduce the 
SFR is a function of scale.  In section 4, we discuss and combine  
 feedbacks from young stellar objects (YSO) and  supernovae (SN)  and  radiation and winds from massive stars.
 We find they can truncate the cascade by the end of  a typical outer scale free fall time, but only after the  cascade survives unfettered to pc scales. 
In section 5, we comment on how the feedback is consistent with 
supplying  $HI$ back to the ISM and  supplying  a substrate of turbulent fluctuations for the next incarnation of fragmentation.
We conclude in section 6.

\section{Mass Conserving Cascade }
\label{sec2}


\subsection{Virial equipartition and basic  framework}
\label{vebf}
Molecular clouds with radial scales $50$pc down
to clump scales $\sim 0.5$pc were long   thought
to obey the Larson scaling relations for velocity dispersion $\sigma \propto  R^{1/2}$
and  density  $\rho(R) \propto R^{-1}$ respectively \citep{Larson1981,Solomon+1987}, 
so that   $M(R) \propto R^2$.
The constants of proportionality were seemingly consistent with virial equipartition  $\alpha_v\sim 1$
where  $\alpha_{v}  \equiv {3GM\over 5\sigma^2 R}$ is a minimalist virial parameter \citep{Bertoldi+1992},  for spherical clouds.  
However, \citet{Heyer+2009} found
that updated measurements are more consistent with 
$\sigma \propto  \Sigma^{1/2} R^{1/2}$ 
and  $\alpha_v <1$, with the surface density $\Sigma$ varying in different regions.  
Others  suggest that that the velocity dispersion may  be primarily infall. \citep{Vaz+2008,Ballesteros+2011,Ballesteros+2012}.  Nearly $90\%$ of the \citet{Heyer+2009} cloud sizes are larger than $1$pc.

 \citet{Field+2011} suggested that the  \citet{Heyer+2009} results may still be consistent
 with clouds in VE  but only if (i) external pressure is taken into account and (ii) clouds are at  a "critical surface density"
     \citep{Chieze1987,Elmegreen1989} 
which minimizes the equlibrium value of  $\sigma^2/R$.
This is  obtained by extremizing the generalized VE equation
 \beq
 {\sigma^2\over R}  ={1\over 3}\left(\pi \Gamma G \Sigma + {4 P_e \over \Sigma}\right), 
\label{sigmac}
 \eeq
 with respect to $\Sigma$  to find $\Sigma_c=13.9 \left(P_e/\Gamma\over 10^5{\rm erg/cm^3}\right)^{1/2}{\ \rm g/cm^2}$, 
 where $P_e$ is the external pressure,  $\Gamma$ accounts for cloud density structure ($\Gamma =3/5$ for uniform density; $\Gamma =0.73$ for isothermal sphere). Plugging $\Sigma_c$ into Eq. (\ref{sigmac})  and solving for $R$ then 
 gives $R=R_c \propto \sigma^2$. Complementarily, replacing $\Sigma$ by $\Sigma_c$  in Eq. (\ref{sigmac})
  and writing $R= (\pi \Sigma_c/M)^{-1/2}$   and solving for $M$ gives $M=M_c \propto \sigma^4 \propto R ^2$. 
  These proprotionalities  also follow if we use $\Sigma \equiv  M/4\pi R^2$ and instead extremize equation (\ref{sigmac}) with respect to 
  $R$.
 For constant  $P_e$  in a given  complex,  Larson-type   scalings  therefore   emerge if clouds are
  at their critical values.
Explaining the full range of clouds requires $P_e$ to vary by 2 orders of magnitude between regions, but gravity from a shell of $\rm HI$ \citep{Elmegreen1989}  is a gravitational alternative to external pressure.

Clouds might naturally appear at their critical values
because  the critical mass  varies as  $\sigma^4$,  and so once the  largest scale cloud forms, 
its fragmentation can only occur after supersonic $\sigma$ decays  enough, likely via shocks,   \citep{Keto+2019} for the
fixed cloud mass to be of order twice the critical mass.   As subunits collapse and
 establish quasi-virial equilibrium, these subunits would themselves be 
 close to the  marginally unstable critical masses for the next stage in the cascade, and so on.  
If fragmentation into two subunits were  somehow to  occur   before a cloud mass were twice the critical mass, then  at least one of the fragments would be rapidly unstable and shorter lasting that a critical cloud of that same mass. Therefore the probability of seeing  clouds near critical   would be  higher than seeing sub or super-critical clouds.
  
  All that said,  the precision to which  VE of clouds can  be  determined from observations  
remains  low \citep{Singh+2019}. Clouds  also deviate from the spherical symmetry commonly assumed. Nevertheless,  we push forward  and assume that  the plausibility arguments above can justify use of observed Larson-type scalings.

Initial fragmentation likely requires that clouds  not only have sufficient gravity 
 but  also  pre-existing density  fluctuations 
because 
 free fall times vary as   $\rho^{-1/2}$. Density fluctuations as high as  $30\%$  over background may be needed to facilitate subunits to collapse before the overall cloud does \citep{Toala+2015}.
Plausibly, these fluctuations can arise  from interacting supersonic velocity fluctuations 
 whose energy is  supplied by a combination of gravity itself, and externally, e.g.. SN driven,  turbulence 
 \citep{MacLow+2004,Kritsuk+2013a,Padoan+2016,Kritsuk+2017}.



 We self-consistently ignore magnetic fields for present purposes on the following grounds:
 magnetic fields in galaxies originate from  kinetic energy from velocity flows.
On the  $\sim 50$pc  scales from which the largest
GMC form and initiate the fragmentation cascade,  
 $\rho v_T^2  \ge \rho v_A^2$.   If  Larson-type relations hold, then $\rho v_T^2 \sim$ constant,  so  this inequality is maintained throughout the  fragmentation inertial range.  
Debate remains \citep[e.g.][]{Federrath2015,Krumholz+2019}, but
observations at least  do not reveal  dynamically dominant magnetic fields above core scales
\citep{Crutcher2012,Nixon+2019}.
The  concepts  herein  apply whether magnetic fields are at or below turbulent equipartition.

\subsection{Mass spectrum 
}




The  total mass contained in clouds of radius $R$ within a  nested complex is \citep{Field+2008} 
\beq
M_R = \int^x_0 M(x'){dN\over dx'}dx' =\int^{M(R)}_0 M'{dN\over dM'}dM'
\label{MR}
\ee
where $x\equiv R/R_L$,   $R_L$  and $M_L=\int^1_0 M(x'){dN\over dx'}dx' $ 
are the  outer  (largest) scale  GMC radius and mass,  $dN/dx'$  and $dN/dM'$ are the number of clouds within the scale range $dx'$  and mass range
$dM'$ respectively.  

Mass conservation though an unfettered cascade 
in which critical  clouds evolve on a dynamical, or approximate free fall time, means
that that ${\dot M}_R \simeq M_R/t_{ff}(R)=$ constant for all $R$.  
Using cloud scaling relations from Sec. \ref{vebf}, we have  $M\propto R^2$ and thus
 \beq
t_{ff}(R)\propto \rho^{-1/2} \propto (R/M(R))^{3/2}\propto R^{1/2}\propto M^{1/4}.
\label{tff}
\eeq
Dividing equation  (\ref{MR}) by $t_{ff} \propto M^{1/4}$
and taking $dM\sim M$ gives  $M^{7/4}  dN/dM =$ constant, or
\beq
dN/dM  \propto M^{-7/4}.
\label{mscale}
\eeq
This is  consistent with the observed range $1.3 \le  d Log N/dLog M \le 1.9$ 
\citep{Myers1983,Casoli+1984,Solomon+1987,Loren1989,Williams+1995,Snell+2002,Blitz+2007}.
Since $M\propto x^2$,  equation (\ref{mscale}) is also equivalent to
\beq
{dN\over dx}\propto x^{-2.5}.
\label{8a}
\eeq

That  observations  down to  clump scales are 
roughly consistent with these spectra and the Larson-type relations 
highlights a consistency with the unfettered mass fragmentation cascade.
   But observations also  require $\ep_{ff}<<1$.
 This can be explained if star-formation is  blocked by cascade-abating feedback only on relatively small
 scales. This explanation  depends on the fact that 
   less feedback momentum  is required to abate the cascade at  small scales 
  than large scales, as discussed in the next section.
  Some mass continues its cascade below the truncation scale
   to supply star formation, and can still  exhibit a similar $dN/dx$ spectra 
  on core scales \citep{Lada+1991}, but with a coefficient that would reflects the volume filling fraction (and thus mass fraction)   that avoided  feedback.






\section{Scale Dependent Momentum and Energy Cascade Rates}

For a mass  conserving fragmentation cascade in each GMC complex, the no-feedback scalar momentum deposition rate on scale $R$ from  collapse is 
\beq
\begin{array}{l}
\frac{{d{P_g}}}{{dt}}(R) =\rho v_{ff}^2\cdot 4\pi R_L^2  x^2\cdot xdN/dx 
\simeq4\pi\rho_L GM_LR_L x^{1/2}\\
\simeq 
3.5 \times {10^{31}} x^{1/2}
\left(\frac{M_L}{10^6{M_\odot}}\right)^2\left(\frac{R_L}{50{\rm{pc}}}\right)^{-2}
\ {\rm{g}}.{\rm{cm/}}{{\rm{s}}^{\rm{2}}},\\
\end{array}\label{mom}
\eeq
where we have used equations (\ref{tff}),  (\ref{8a})
${M_L} = {10^6}{M_ \odot },{R_L} = 50{\rm{pc}}$, and the
scale independence of $\rho v_{ff}^2$ implied by Larson's relations
 to replace this latter quantity  by $\rho_L v_{ff,L}^2$, where $v_{ff}$ and $v_{ff,L}$
 are the free fall speeds for arbitrary scale $R$ and  scale $R_L$ respectively.  
Similarly, the energy deposition rate  is 
\beq
\begin{array}{l}
\frac{{d{E_g}}}{{dt}}(R) =\rho  v_{ff}^3\cdot 4\pi R_L^2x^2\cdot xdN/dx\\
\simeq 4\pi \rho_L G^{3/2}M_L^{3/2}R_L^{1/2} x\\
\simeq 
3.3 \times {10^{37}}x
\left(\frac{M_L}{10^6{M_\odot}}\right)^{5/2}\left(\frac{R_L}{50{\rm{pc}}}\right)^{-1.5}
 {\rm{g}}.{\rm{cm^2/}}{{\rm{s}}^{\rm{3}}},
\end{array}
\label{energy}\eeq
where have  also used $v_{ff}\propto R^{-1/2}$.
Equations (\ref{mom}) and (\ref{energy})
show that the free fall energy and momentum feedback requirements  to abate the cascade and reduce SFR are less demanding on smaller scales.
Abating the free fall momentum (energy) flux is the  more (less)  stringent requirement if whatever supplies the  feedback has a larger (smaller) speed than  free fall  on the feedback  coupling scale.  Below we focus on  momentum
feedback.



 
\section{ Transition Scale and Feedback}



From equation (\ref{mom}),  the required momentum for abating a fragmentation cascade at $R\sim 0.5$pc for example, is 10 times smaller than at $R=R_L=50$pc. 
Feedback that  can only truncate the fragmentation cascade  after it proceeds to these small scales would  not  threaten unfettered  Larson-type relations at larger scales.
Feedback from low mass stars is dominated by YSOs, and feedback from massive stars  comes from SN,  winds, and radiation
\citep{Krumholz+2014,Kim+2015,Iffrig+2015, Kim+2019,Grudic+2019}.
 We consider each in turn below.
 Given these feedback sources, we then  determine the maximum transition scale 
 below  which the fragmentation is  abated.

 \subsection{YSO Feedback}
  \cite{Matzner+2000}  found theoretically  that YSOs  can deliver  $\sim 40$km/s of momentum per unit mass of stars formed. 
  Simulations   of  \cite{Carroll+2009}   are consistent with  similar values.   
 For one of  our 2000  outer scale GMC, the SFR  is $2\times 10^{-3}\sim M_{\odot}/{\rm yr}$    over  its  $5.5\times 10^6$yr free fall time, so there is then 
 \beq
P_{Y}=5.5\times 10^6 \cdot 2 \times 10^{-3} M_\odot \cdot 40{\rm km/s} = 
8.8 \times 10^{43}{\rm g . cm/s }
\label{12}
 \eeq
 of momentum delivered by YSOs.
 The latter is also consistent with 
 observational   studies  \citep{Quillen+2005,Brunt+2009,Nakamura+2011} suggesting that  stellar outflows can inject a   turbulent velocity dispersion for typical molecular clouds
   $\sim 1.5$ km/s  at $10\%$ volume filling fraction at $\sim 0.2$pc scales. 
\footnote{Observationally, outflow feedback  may be  underestimated by principal component analysis  (PCA): \cite{Carroll+2010} showed that  PCA,   when  used to  identify turbulent  driving scales \citep{Brunt+2009}, is biased toward low amplitude large-scale velocity structures, even if energetically subdominant.} 

 \subsection{SN Feedback}
For a $10^6M_\odot$ GMC, and a stellar mass spectrum of $dN/dm \propto m^{-2.3}$ 
 with stellar mass range $0.1 \le m \equiv M_* /M_\odot \le 100$ \citep{Chabrier2003},
 there would be 
\beq
N(m_0)=  N_0 \int_{m_0}^{100} m^{-2.3} dm, 
\label{13}
\eeq
 stars of $m>m_0$ producing SN
 during a time scale $t_{ff,L}$,  where $N_0=1721$ is the normalization that   ensures  
 $N_0 M_\odot \int_{m_0}^{100} m^{-1.3} dm=10^4$ or 
  $ 10^6(\ep_{ff}/0.01)\sim 10^4$ for a free fall time $t_{ff,L}$.
Since only stars  with  $m_0\gsim 40$ could contribute over this time scale \citep{Maeder+1987},
we obtain $N(m_0=40)\sim 7$.  These  SN  would inject  an initial momentum
$P_{SN,I}(M_*)\simeq 
7 \sqrt {\left(2 {M_*\over 2 M_\odot}{E\over 10^{51}{\rm erg/s}}\right)}=6.3 \times 10^{43}$cm/s,  for $M_*=40 M_\odot$ (assuming ejected mass $\sim M_*/2$ for $M_*>>M_\odot$)
and  $E= 10^{51}$erg.
But  during  the Sedov phase, approximately $\sim 8$ times more momentum is injected 
from the work done by the post shock gas \citep{Iffrig+2015,Kim+2015,Martizzi+2016}.
We therefore obtain the momentum contribution
\beq
P_{SN}\simeq 8 P_{SN,I}(40M_\odot)=5 \times 10^{44}{\rm g\cdot cm /s}.
\label{14}
\eeq

There is also a contribution from background SN: 
most  Galactic  SN 
 occur within  the inner $5$kpc  thin disc, of  thickness of $\sim 300$pc
or   volume   $2.4 \times 10^{10}{\rm pc^3}$.  Since our  
 2000,  $50$pc outer scale GMC's have volume 
 $\simeq 10^9{\rm pc^3}$,
 they fill just  4\% of the  total region.
 If we put $m_0\sim 8$ in equation (\ref{13}), we  obtain the total of $32 >>2$ stars that produce SN
 from a given cloud complex, even if that complex has  disappeared. 
 Assuming that the complexes do not last more than a a time scale $t_{ff,L}$
 (justified later),  then most  SN will  not occur in existing complexes.
 However,  averaged over time scales $>> 10$Myr, 
some of these   SN  will be randomly  cospatial with  existing GMC.
For SN rate $1/40 {\rm yr}$ \citep{Tammann+1994,Li+2011}
there would be roughly $ (1 /40{\rm yr} )( 0.04/2000)( 5.5 \times 10^6{\rm yr})\sim
3$ such background SN per complex free-fall time contributing to the feedback. Their
momentum contribution is  
\beq
P_{SN,b}\sim 3 P_{SN}(10M_\odot) 
=1.4\times 10^{44}{\rm g\cdot cm/s}.
\label{15}
\eeq

 \subsection{Radiation Feedback}
 
Theory and simulation conclude that   per unit mass of star formation, radiation supplies   $\sim 200$km/s  of momentum \citep[e.g][]{Kim+2019},  beginning  shortly after the first stars of a complex form and mostly due to ionizing radiation.
The consequent momentum deposited  in a complex free fall time is then
   \beq
    P_{R}=5.5\times 10^6 \cdot 2 \times 10^{-3} M_\odot \cdot 200{\rm km/s} = 
4.4 \times 10^{44}{\rm g . cm/s }.
\label{16}
\eeq

 \subsection{Winds from Massive Stars}
The momentum imparted per star  from massive stellar winds
is  $P_{w,*}={M_e} v_\infty\le {M_*} v_\infty$   where $M_e( M_*)$ is the mass ejected
over the entire the active wind before SN and  $v_\infty\lsim 2.5 v_{es}=3\sqrt (GM_*/R)^{1/2}$ 
is the terminal wind speed \citep{Kudritzki+2000}.
Using $R/R_\odot\sim 1.3(M_*/M_\odot)^{3/5}$,  
\citep{Demrican+1991}
then $v_{es}(m)\simeq 5.36 \times 10^7 m^{1/5} $cm/s.
For the total momentum contribution integrated over the mass function of massive stars,
we    divide the contribution  into (i) the most massive stars that  eject $\sim 1/2$ or more  their mass and $v_\infty=2.5 v_{esc}$, and (ii) slightly lower mass stars which eject  $\sim 0.1$ of their mass with $v_\infty\sim 1.5 v_{esc}$ during the wind loss phase.
We then have
\beq
\begin{array}{l}
P_{W}\simeq \ep_WN_0 M_\odot\left[{1.25} \int_{20}^{100}  m^{-1.3}  v_{es} dm
+\ 0.15 \int_{10}^{20} m^{-1.3}  v_{es} dm \right]
\\=4.1 (\ep_W/0.15)\times 10^{43} {\rm cm.g/s},
\end{array}
\label{wind}
\eeq
where $\ep_W$ is an efficiency factor.
That equation (\ref{wind})  with $\ep_W=1$ is comparable to the contrinution from SN 
is consistent with previous   detailed work  \citep[e.g.][]{Fierlinger+2016}, 
but  is  likely an over-estimate because   hot gas from winds may escape  an inhomogeneous
environment: \cite{Rosen+2014} estimate $\ep_W\sim 0.15$.

 \subsection{Total Feedback and the Transition Scale}
 
Combining equations (\ref{12}) (\ref{14}) (\ref{15}) , (\ref{16}), and (\ref{wind})
gives
 the total  momentum delivered over an outer GMC  free-fall  $t_{ff,L}$ as
 \beq
 P_{T}=P_{Y} +P_{SN}(M_{*0})+P_{SN,b}+ P_{R}+P_W = 1.2 \times 10^{45}{\rm erg /s}.
 \label{ptot}
 \eeq
 Setting this equal to equation (\ref{mom}) multiplied by  $t_{ff,L}=5.5$Myr gives
 the  scale of equality of   $x=0.044$, or $R=2.2$pc.
Thus, the combination of these  momentum feedbacks during a time $t_{ff,L}$
could truncate the cascade and  dissipate the complex below these scales.
 This  is  qualitatively consistent with the  destruction seen in simulation movies \citep[e.g][]{Grudic+2019}).
The trickle of star formation
arises gas that does make it through to smaller scales. The mass spectrum for  this subset of mass that avoids the feedback ad continues to fragment
 may remain
similar to that above the transition scale, but normalized by a much smaller filling fraction.

During the time scale of $t_{ff,L}$ the 7 SN from the complex cascade and the 3 background SN
would inject $\sim 10^{52}$erg into the complex. Ionizing radiation would inject photons at rate of 
$\sim 10^{50}/s$ \citep{McKee+1997} for an energy input $> 3.6 \times 10^{53}$erg.
The amount of energy needed to unbind all the $H_2$ in the gas  of the complex is about
$9\times 10^{52}$ erg, so these sources do have enough energy to unbind the molecules
at $3\%$ efficiency.  Thus, once the cascade is abated,  much of the gas may  be returned to the ISM as HI.


\section{Reseeding Turbulence and HI}

SN   are  a source of galactic turbulence for much of the ISM  \citep{Spitzer1978,Norman+1996}, supplemented by a substrate of 
shear driven turbulence \citep{Sellwood+1999}.
This  also seeds turbulence in molecular clouds  \citep{MacLow+2004,Padoan+2016}. 
The latter is important, as discussed in section \ref{sec2}, since a source of subunit density fluctuations  within a parent cloud is needed to seed subunit fragmentation.
In addition, feedback from  SN and  radiation   can convert  gas into HII which cools into HI.  This   can also account for the observed mass  of HI in shells,  energy content of shells and supershells,  and depletion of  $\rm H_2$ between
spiral arms. We discuss each below.

 \cite{Heiles1979}  found that most of the atomic hydrogen is contained within  45  shells (defined by energy content  $< 3 \ts 10^{52}$ erg) and 18 supershells with larger
energies. The latter dominate both the shell 
masses and  energies  by factors of 10 to 100.
The total HI mass in shells is about $M_{HI}\sim 4 \ts 10^8M_\odot$.
\footnote{The total HI mass in the Galaxy is closer to  $ 4 \ts 10^9M_\odot$
\citep{Dickey+1990,Sparke+2006}}
\cite{Heiles1979} estimated a typical age of $t_{sb}\sim 10^7$yr
for the lifetime of  HI supershells,
so  the  needed rate of HI  
supply would be $\sim M_{HI}/t_{sb}=40M_\odot$/yr. 
Dividing the typical SN injected momentum of $\sim 5.1 \times 10^{43}{\rm g\cdot cm/s}$ 
(for a $10M_\odot$ progenitor including the  factor of $8$ discussed in the previous section)
by typical flow speeds $\sigma_{SN}\sim 10$km/s when  remnants break up
gives a mass influenced per SN of $M\sim 2.5\times 10^4M_\odot$.
Multiplying  by the galactic SN rate  1/40yr, then gives $625M_\odot$/yr,
so only $6.4\%$  of this   need be involved to account for the  observations even if we
just appeal  to SN. 
Dviding the  total  kinetic energy  $\sim 2\ts 10^{54}$erg in observed shells 
\citep{Heiles1979} by typical  lifetimes $t_{sb}\sim 10^7$yr, gives a required energy injection rate $6.7\times 10^{39}$erg/s, which can also be accomplished  by SN  during that time scale, which inject
$10^{51} \times 0.02/{\rm yr} /3\times 10^7{\rm sec /yr}=6.6 \times 10^{41}$ergs.



Finally
note that 
the bulk of the total $\sim 2 \ts 10^9 M_\odot$ of  $\rm H_2$ in the Galaxy resides in spiral arms. The absence of $\rm H_2$ in the 
inter-arm regions means that the molecular gas
must be converted into $\rm HI$ by the time 
that  material moves through the arms.
At radii of 5kpc, there  are $\sim 4$ arms \citep{Vallee2005}
each of which take $\sim 1.3 \times 10^8$yr to pass through as
the gas motion has a significant component along the arms. 
This lengthens the time scale for gas to pass through the arms 
from the value of  $\sim 1/4$ of a rotation period  
for  purely radial arms.
The  minimum rate of $\rm H I$ formation needed to account for the 
absence of $\rm {H_2}$ in the inter-arm regions is  then
 $\sim {2\ts 10^{9}M_\odot\over  1.3 \times 10^8{\rm yr}}
\sim  15.4  M_\odot$/yr. 
But since feedback destroys the complex on an  outer overturn time,
the supply rate can be as high as  $\sim {2\ts 10^{9}M_\odot\over  5.5 \times 10^6{\rm yr}}
\sim  360 M_\odot$/yr so this is  fast enough. 


\section{Conclusions}

We have discussed how molecular cloud complexes, interpreted as  mass-conserving fragmentation cascades, 
have properties   roughly consistent with  an unfettered cascade with clouds near their critical unstable mass
over most of their inertial range, while still producing star formation with low efficiency.  This can work because  the
feedback momentum needed to  abate the cascade and  destroy the cloud complex decreases with decreasing
scale. The primary sources of feedback momentum available from YSOs, and massive stars
provide enough momentum and energy to abate the cascade on a free-fall time of the outer complex scale, and dissipate
the complex, but only after the fragmentation cascade
reaches pc scales. In this way, the star formation is damped while  the super-transition scale inertial range retains Larson-type relations and an unfettered size spectrum power law consistent with observations. 
Below the transition scale, there will be some mass that slips through the feedback and this goes onto form stars.
For that gas, the number-size spectrum may be retained, but with a normalization coefficient that is reduced to reflect
participation of only the mass fraction that beats the feedback.

Pinning down the existence and value of the transition scale, or using the framework as a guide to help interpret 
observations may be useful.  The   complexity of the interaction between feedback, fragmentation, and turbulence has and continues to warrant more detailed calculations and simulations, 
but  the goal of  synthesizing various   pieces into an overall framework 
remains a conceptually important complement to improved understanding of the microphysics.





\section*{Acknowledgments} 
Thanks to  G. Field for related discussions on clouds, supershells, and HI, and to E.Keto  for  particuarly useful comments on the manuscript. I acknowledge support from NSF Grant  AST-1813298, KITP (UC Santa Barbara) funded by  NSF Grant PHY-1748958, and the Aspen Center for Physics funded by NSF Grant PHY-1607611.



\end{document}